\documentclass[12pt]{iopart}
\usepackage{amssymb,amsthm,bm}
\usepackage[colorlinks=true, allcolors=blue]{hyperref}

\newtheorem{theorem}{Theorem}

\begin{document}
\title{Generalized Euler angles for a unitary control of the Hamiltonian system}
\author{Seungjin Lee, Kyunghyun Baek and Jeongho Bang}
\address{Electronics and Telecommunications Research Institute, \protect\\ Gajeong-ro 218, Daejeon, Republic of Korea, 34129}
\ead{seungjin.lee@etri.re.kr}
\begin{abstract}
    We provide an angular parametrization of the special unitary group $\textrm{SU}(2^{n})$ generalizing Euler angles for $\textrm{SU}(2)$ by successively applying the KAK decomposition. We then determine constraint equations for the parametric curve of generalized Euler angles corresponding to the exponential curve of a given Hamiltonian. The constraint equations are in the form of first-order differential-algebraic equations and resemble Wei-Norman equations of canonical coordinates of the second kind for $\textrm{SU}(2^{n})$.
\end{abstract}

\noindent{\it Keywords\/}: Hamiltonian simulation, uniform finite generation, KAK decomposition, Euler angles, differential-algebraic equations


\section{Introduction}
\label{sec:intro}

The simulation of a quantum system through a controllable quantum system, often called the Hamiltonian simulation, has been a prominent application in quantum computation. Along with the vitality of the Hamiltonian simulation for diverse target systems, the particular interest in the Hamiltonian simulation resides in its anticipated computational power over the classical implementation owing to the intrinsic quantum nature of the control system.

In the Hamiltonian simulation, it is essential to establish the connection between the control parameters of the control system to the unitary evolution generated by the Hamiltonian of the target system embedded in the control system. This necessitates a comprehensive specification of the control system, which includes the complete set of control parameters and the relationship between a particular Hamiltonian and control parameters.

In mathematical terms, the unitarity of a quantum system's evolution indicates that the controllability of the control system for a Hamiltonian simulation amounts to a parametrization of the unitary group embedding the evolution of the target system. For instance, given a two-level quantum system, an evolution of the system can be identified with an element of $\textrm{U}(2)$. Therefore, up to the overall phase factor, a conventional Euler angle parametrization of $\textrm{SU}(2)$ provides a set of control parameters along with the polar decomposition.

A particularly practical class of parametrizations involves decomposing the unitary group into one-parameter subgroups. With this approach, an element of the unitary group can be expressed as a finite string of elements of subgroups in which the order of subgroups in the string is fixed and independent of the decomposed element. Each parameter in the string provides complete control over the corresponding subgroup, allowing for precise manipulation of the target element. Such decomposition is often called the uniform finite generation of the unitary group and has acquired a particular interest since the group-generating string can be interpreted as a sequential implementation of elements forming the string.

The uniform finite generation of the unitary group has been extensively studied in the context of the controllability of systems on Lie groups (see, for instance, \cite{dalessandro2001} and related references). The first existence of a uniform finite generation has been established in \cite{jurdjevic1972}, and various explicit constructions have been presented in \cite{dalessandro2001,lowenthal1972,divakaran1980}. 

The KAK decomposition of a compact Lie group has provided systematic constructions of the uniform finite generation of the unitary group and has been studied in \cite{crouch1983,leite1991,tilma2002,khaneja2001,bullock2004a,dalessandro2007}. In particular, in \cite{khaneja2001}, the KAK decomposition of $\textrm{SU}(2^{n})$ has been applied to obtain a uniform finite generation of $\textrm{SU}(2^{n})$ which is relevant to unitary evolutions embedded in the control system consisting of $n$ two-level (or qubit) systems.

However, despite diverse constructions of the uniform finite generation of the unitary group, the direct interconnection between the Hamiltonian simulation and the uniform finite generation has yet to be extensively studied. In particular, the relation between the integral curve generated by the given Hamiltonian of the system via the exponential map and the parametric curve for the control parameters has not been explicitly established apart from the well-known equations for Euler angles \cite{goldstein2002}.

In this work, we provide a construction of the uniform finite generation of the unitary group $\textrm{SU}(2^{n})$ by extending results in \cite{khaneja2001,bullock2004a,dalessandro2007,hai-rui2008}. Having a uniform finite generation of $\textrm{SU}(2^{n})$, we establish a set of constraint equations for the parametric curve corresponding to the integral curve generated by the Hamiltonian of the target system via the exponential map, which provides the direct interconnection between the Hamiltonian simulation and the uniform finite generation of the $n$-qubit system.

The paper is organized as follows. In the next section, we provide a decomposition of the unitary group $\textrm{SU}(2^{n})$ by successively applying the KAK decomposition. The decomposition can be considered a refinement of works in \cite{khaneja2001,hai-rui2008}, having a particular intention to systematically relate the decomposition with the Hamiltonian simulation. 

We then show that for a unitary evolution of a Hamiltonian system, one can obtain a system of differential-algebraic equations (DAEs) for the parametric curve identical to the unitary evolution. The construction of the DAEs resembles that of the Wei-Norman equations in \cite{wei1964,owren2001,charzyski2013}, whose parametrization is the set of canonical coordinates of the second kind. The last section is devoted to the summary and outlook, discussing possible research directions for our work.

\section{KAK decomposition and generalized Euler angles for $\textrm{SU}(2^{n})$}
\label{sec:KAK}

An evolution of a Hamiltonian system embedded in an $n$-qubit system can be represented as an element in $\textrm{SU}(2^{n})$ by ignoring the overall phase factor. Therefore, a uniform finite generation of $\textrm{SU}(2^{n})$ provides a set of control parameters for the evolution of the system.

The KAK decomposition of the unitary group provides a systematic construction of a uniform finite generation of the unitary group. The KAK decomposition in the context of a unitary control of a Hamiltonian system has been considered in \cite{khaneja2001}. In this section, we extend the method in \cite{khaneja2001} to decompose $\textrm{SU}(2^{n})$ in a recursive manner, which is somewhat close to the exposition in \cite{hai-rui2008}.

A decomposition of the KAK type for a compact connected semisimple Lie group $G$ is associated with an orthogonal symmetric Lie algebra of the compact type of the corresponding Lie algebra $\mathfrak{g}$. For a compact Lie algebra, the construction of an orthogonal symmetric Lie algebra amounts to a vector space decomposition $\mathfrak{g} = \mathfrak{k} \oplus \mathfrak{p}$ such that \cite{helgason1979}
\begin{eqnarray}
	\left[ \mathfrak{k} \mathfrak{k} \right] \subset \mathfrak{k},\quad \left[ \mathfrak{k} \mathfrak{p} \right] = \mathfrak{p}, \quad \left[ \mathfrak{p} \mathfrak{p} \right] \subset \mathfrak{k}.
\end{eqnarray}

Given the orthogonal symmetric Lie algebra $(\mathfrak{g}, \mathfrak{k})$ we have a KAK decomposition of $G$ as an application of the following theorem whose proof can be found in \cite{helgason1979}:
\begin{theorem}[KAK decomposition]
	Let $G$ be a connected compact Lie group whose Lie algebra $\mathfrak{g}$ is semisimple, and $(\mathfrak{g},\mathfrak{k})$ be an orthogonal symmetric Lie algebra of the compact type. Then for  all $g \in G$, there exist $k_{1}, k_{2} \in \mathfrak{k}$ and $a \in \mathfrak{a}$ such that $g = \exp\left( k_{1} \right) \exp\left( a \right) \exp\left( k_{2} \right)$ in which $\mathfrak{a}$ is the maximal abelian subalgebra of $\mathfrak{p}$.
\end{theorem}

In order to construct an orthogonal symmetric Lie algebra for the Lie algebra $\mathfrak{su}(2^{n})$ of $\textrm{SU}(2^{n})$, it is somewhat convenient to consider the tensor product construction of $\mathfrak{su}(2^{n})$, which manifestly exhibits the composite nature of the $n$-qubit system. For the tensor product construction, we mean an associative algebra recursively constructed by
\begin{eqnarray}
	\mathfrak{u}(2) =& \mathbb{R}\left\{ \sigma_{a}: a = 0,1,2,3 \right\}\\
	\mathfrak{u}(2^{n}) =& i\sigma_{a} \otimes \mathfrak{u}(2^{n-1})
\end{eqnarray}
where $\sigma_{a} = \left( -\frac{i}{2} \mathbf{1}, \frac{1}{2} \mathbf{i}, \frac{1}{2} \mathbf{j}, \frac{1}{2} \mathbf{k} \right)$ with the following multiplication table of quaternions $\{\mathbf{1}, \mathbf{i}, \mathbf{j}, \mathbf{k}\}$:
\begin{eqnarray}
	\begin{array}{c|cccc}
		&\mathbf{1} & \mathbf{i} & \mathbf{j} & \mathbf{k} \\
		\hline
		\mathbf{1} & \mathbf{1} & \mathbf{i} & \mathbf{j} & \mathbf{k}\\
		\mathbf{i} & \mathbf{i} & - \mathbf{1} & \mathbf{k} & -\mathbf{j} \\
		\mathbf{j} & \mathbf{j} & - \mathbf{k} & - \mathbf{1} & \mathbf{i} \\
		\mathbf{k} & \mathbf{k} & \mathbf{j} & -\mathbf{i} &  - \mathbf{1}
	\end{array}.
\end{eqnarray}
One can easily notice that $\mathfrak{u}(2^{n})$ induces the unitary Lie algebra of rank $2^{n}$, so we have $\mathfrak{su}(2^{n})$ by excluding the center of $\mathfrak{u}(2^{n})$.

Given $\mathfrak{su}(2^{n})$ as in the above, we have an orthogonal symmetric Lie algebra $(\mathfrak{su}(2^{n}), \mathfrak{k}_{n})$ constructed by \cite{khaneja2001}
\begin{eqnarray}
	\mathfrak{k}_{n} = \left( i\sigma_{0} \otimes \mathfrak{su}(2^{n-1}) \right) \oplus \left( i\sigma_{3} \otimes \mathfrak{u}(2^{n-1}) \right), \quad \mathfrak{p}_{n} = \bigoplus_{i = 1,2} \left( i\sigma_{i} \otimes \mathfrak{u}(2^{n-1}) \right).
\end{eqnarray}
Consequently, the corresponding KAK decomposition of $\mathfrak{su}(2^{n})$ can be casted into the form of
\begin{eqnarray}
	\textrm{SU}(2^{n}) = \exp(\mathfrak{k}_{n}) \exp(\mathfrak{a}_{n}) \exp(\mathfrak{k}_{n})	
\end{eqnarray}
with $\mathfrak{a}_{n} = i\sigma_{1} \otimes \mathfrak{t}_{n-1}$ in which $\mathfrak{t}_{n}$ denotes the maximal abelian subalgebra of $\mathfrak{u}(2^{n})$.

One can further notice that the corresponding Lie group of $\mathfrak{k}_{n}$ is homeomorphic to $\textrm{U}(1) \times \textrm{SU}(2^{n-1}) \times \textrm{SU}(2^{n-1})$ so that one can apply another KAK decomposition for $\exp(\mathfrak{k}_{n})$ as \cite{hai-rui2008}
\begin{eqnarray}
	\exp(\mathfrak{k}_{n}) = \exp(\mathfrak{k}_{n}') \exp(\mathfrak{a_{n}'}) \exp(\mathfrak{k}_{n}').
\end{eqnarray}
Explicitly, the decomposition of $\textrm{SU}(2^{n-1}) \times \textrm{SU}(2^{n-1})$ is associated with an orthogonal symmetric Lie algebra $\left( \mathfrak{k}_{n}, \mathfrak{k}_{n}' \right)$ constructed by
\begin{eqnarray}
	\mathfrak{k}_{n}' = i\sigma_{0} \otimes \mathfrak{su}(2^{n-1}), \quad \mathfrak{p}_{n}' = i\sigma_{3} \otimes \mathfrak{su}(2^{n-1}), \quad \mathfrak{a}_{n}' = i\sigma_{3} \otimes \mathfrak{t}_{n-1}.
\end{eqnarray}

Altogether, we have a sub-extended KAK decomposition of $\textrm{SU}(2^{n})$ as
\begin{eqnarray}\label{eq:kak}
	\textrm{SU}(2^{n}) = \exp(\mathfrak{k}_{n}') \exp(\mathfrak{a}_{n}') \exp(\mathfrak{k}_{n}') \exp(\mathfrak{a}_{n}) \exp(\mathfrak{k}_{n}') \exp(\mathfrak{a}_{n}') \exp(\mathfrak{k}_{n}').
\end{eqnarray}
By noticing that $\exp(\mathfrak{k}_{n}')$ is isomorphic to $\textrm{SU}(2^{n-1})$ the decomposition in \eref{eq:kak} allows one to recursively decompose $\textrm{SU}(2^{n})$ until we reach $\textrm{SU}(2)$ decomposed into 
\begin{eqnarray}
\exp(\mathbb{R} \sigma_{3}) \exp(\mathbb{R}\sigma_{1}) \exp(\mathbb{R}\sigma_{3}).
\end{eqnarray}
For instance, for $\textrm{SU}(4)$ we have the following decomposition: 
\begin{eqnarray}
	\textrm{SU}(4) = \exp(\mathfrak{k}_{2}') \exp(\mathfrak{a}_{2}') \exp(\mathfrak{k}_{2}') \exp(\mathfrak{a}_{2}) \exp(\mathfrak{k}_{2}') \exp(\mathfrak{a}_{2}') \exp(\mathfrak{k}_{2}')
\end{eqnarray}
in which
\begin{eqnarray} \label{eq:su4-kak-gen}
	e^{\mathfrak{a}_{2}} = e^{ \mathbb{R} \sigma_{10} } e^{ \mathbb{R} \sigma_{13} }, \quad 
	e^{\mathfrak{k}_{2}'} = e^{ \mathbb{R} \sigma_{03} } e^{ \mathbb{R} \sigma_{01} } e^{ \mathbb{R} \sigma_{03} }, \quad
	e^{\mathfrak{a}_{2}'} = e^{ \mathbb{R} \sigma_{30} } e^{ \mathbb{R} \sigma_{33} }
\end{eqnarray}
where $\sigma_{ab} \propto i\sigma_{a} \otimes \sigma_{b}$.

Generators of the decomposition of $\textrm{SU}(4)$ in \eref{eq:su4-kak-gen} illustrate that each one-parameter subgroup of the decomposition in \eref{eq:kak} is generated by a Pauli string in the form of $\sigma_{a_{1}} \otimes \ldots \otimes \sigma_{a_{n}}$. In turn, the space $\Theta_{n}$ of parameters of one-parameter subgroups is homeomorphic to the $l_{n}$-torus
\begin{eqnarray}
	T^{l_{n}} = \underbrace{S^{1} \times \ldots \times S^{1}}_{l_{n}} 
\end{eqnarray}
where $l_{n}$ is the length of the decomposition. At $n=1$, $\Theta_{1}$ corresponds to the space of Euler angles, $\Theta_{n}$ can be thus taken as a generalization of the space of Euler angles. We also note that the length $l_{n}$ of the string in \eref{eq:kak} is given recursively by
\begin{eqnarray}
	l_{1} =3, \quad l_{n} = 4 \times l_{n-1} + 3(n-1).
\end{eqnarray}

\section{Constraint equations for generalized Euler angles}
\label{sec:control}
 
In the previous section, we have shown that up to the overall phase factor, $\textrm{SU}(2^{n})$ can be decomposed into a finite string of one-parameter subgroups. Therefore, given a Hamiltonian, the evolution of the Hamiltonian system can be decomposed as
\begin{eqnarray}\label{eq:control}
	\exp(t H) = e^{k'_{1}(t)}  e^{a'_{1}(t)} e^{k'_{2}(t)} e^{a(t)} e^{k'_{3}(t)} e^{a'_{2}(t)} e^{k'_{4}(t)}.
\end{eqnarray}
for $k'_{1,2,3,4} \in \mathfrak{k}_{n}'$, $a'_{1,2} \in \mathfrak{a}_{n}'$ and $a \in \mathfrak{a}_{n}$. 

As we have discussed in the below of \eref{eq:kak}, $\exp(\textrm{ad}\,\mathfrak{k}'_{n})$ can be recursively decomposed into one-parameter subgroups generated by $\sigma_{a_{1} \ldots a_{n}} \propto i\sigma_{a_{1}} \otimes \sigma_{a_{2} \ldots \sigma_{n}}$ for $a_{i} = 0,1,2,3$. In turn, one obtains
\begin{eqnarray}
	\exp(tH) = e^{\theta_{1}(t) X_{1}} e^{\theta_{2}(t) X_{2}} \ldots e^{\theta_{l_{n}}(t) X_{l_{n}}}
\end{eqnarray}
where $X_{i},\ i = 1, 2, \ldots, l_{n}$ are Pauli strings.

By identifying $\exp(tH)$ with an integral curve generated by a left-invariant vector field $H$, one obtains the following identity by comparing the left-invariant vector fields of both sides in \eref{eq:control} \cite{knapp2013}: 
\begin{eqnarray}
	H = \sum_{i =1}^{l_{n}} \dot{\theta}_{i} \left[ \prod_{j = l_{n}}^{i+1} \exp(- \theta_{j} \textrm{ad}\,X_{j}) \right] X_{i}, \quad \dot{\theta}_{i} = \frac{d \theta_{i}}{dt}
\end{eqnarray}
where $\left( \textrm{ad}\,X \right) Y = \left[ X, Y \right]$ for $X,Y$ in $\mathfrak{su}(2^{n})$. Upon an expansion under a basis of $\mathfrak{su}(2^{n})$, we obtain a system of first-order differential-algebraic equations (DAEs)
\begin{eqnarray} \label{eq:odegeneral}
	h_{k} = \sum_{i =1}^{l_{n}} \dot{\theta}_{i} \left[ \prod_{j = l_{n}}^{i+1} \exp(- \theta_{j} \textrm{ad}\,X_{j}) X_{i} \right]_{k},\quad k = 1, \ldots, 2^{2n} - 1
\end{eqnarray}
where $h_{k}$ and $\left[ \ldots \right]_{k}$ denote the linear coefficients of the $k$-th element of the basis for $H$ and $\left[ \ldots \right]$ respectively. DAEs in \eref{eq:odegeneral} resemble DAEs in \cite{wei1964,owren2001,charzyski2013} in the spirit of comparing left-invariant vector fields but differ in the sense that in \cite{wei1964,owren2001,charzyski2013} the decomposition corresponds to canonical coordinates of the second kind.

If we fix $\sigma_{a_{1} \ldots a_{n}} = 2i \sigma_{a_{1}} \otimes \sigma_{a_{2} \ldots a_{n}}$, we have
\begin{eqnarray}
	\left[ \sigma_{a_{1} \ldots a_{n}}, \left[ \sigma_{a_{1} \ldots a_{n}}, \sigma_{b_{1} \ldots b_{n}} \right] \right]
	=
	\cases{
		0 &if $\left[ \sigma_{a_{1}\ldots a_{n}}, \sigma_{b_{1} \ldots b_{n}} \right] = 0$,\\
		- \sigma_{b_{1} \ldots b_{n}} \quad &otherwise,
	}
\end{eqnarray}
so that
\begin{eqnarray} \label{eq:cyclic}
	\exp(\textrm{ad}\, \theta \sigma_{a_{1}\ldots a_{n}}) \sigma_{b_{1} \ldots b_{n}} \nonumber \\ = 
	\cases{
		\sigma_{b_{1} \ldots b_{n}} &if $\left[ \sigma_{a_{1}\ldots a_{n}}, \sigma_{b_{1} \ldots b_{n}} \right] = 0$,\\
		\cos\theta \sigma_{b_{1} \ldots b_{n}} + \sin\theta \left[ \sigma_{a_{1} \ldots a_{n}}, \sigma_{b_{1} \ldots b_{n}} \right] \quad &otherwise.  
	}
\end{eqnarray}
Consequently, the action of $\exp(\textrm{ad}\, \theta \sigma_{a_{1} \ldots a_{n}})$ always decomposes $\mathfrak{su}(2^{n})$ into
\begin{eqnarray}
	\mathfrak{su}(2^{n}) = V_{1} \oplus V_{2} \oplus \ldots V_{k} \oplus W_{2^{2n} -2k - 1}
\end{eqnarray}
on which $W_{2^{2n} - 2k - 1}$ is an $2^{2n} - 2k -1$-dimensional invariant subspace and 
\begin{eqnarray}
	\exp(\textrm{ad}\, \theta \sigma_{a_{1} \ldots a_{n}}) \vert_{V_{i}}
	= 
	\left(
	\begin{array}{cc}
		\cos \theta & \sin \theta \\
		-\sin \theta & \cos \theta
	\end{array} \right), \quad i = 1,2,\ldots, k.
\end{eqnarray}
By combining \eref{eq:cyclic} with \eref{eq:odegeneral} DAEs in \eref{eq:odegeneral} has the form of
\begin{eqnarray} \label{eq:dae-matrix-form}
	\bm{h} = \bm{J}_{n} . \dot{\bm{\theta}},
\end{eqnarray}
where $\bm{J}_{n}$ is a $\left( 2^{2n} -1 \right) \times l_{n} $ matrix depending on $\left(\cos \theta_{i},\ \sin \theta_{i} \right),\ i = 1, 2, \ldots, l_{n}$ and
\begin{eqnarray}
	\bm{h} = \left( h_{1}, h_{2}, \ldots, h_{2^{2n}-1} \right)^{T}, \quad \dot{\bm{\theta}} = \left( \dot{\theta}_{1}, \dot{\theta}_{2}, \ldots, \dot{\theta}_{l_{n}} \right)^{T}.
\end{eqnarray}

For $\mathfrak{su}(2)$, DAEs in \eref{eq:odegeneral} can be casted into the form of
\begin{eqnarray}
	\begin{array}{l}
		h_{1} = \sin\beta \sin\gamma \dot{\alpha} + \cos\gamma \dot{\beta}\\
		h_{2} = \sin\beta \cos\gamma \dot{\alpha} - \sin\gamma \dot{\beta}\label{eq:eulerangluar}\\
		h_{3} = \cos\beta \dot{\alpha} + \dot{\gamma}
	\end{array}
\end{eqnarray}
or equivalently
\begin{eqnarray}
	\left(
	\begin{array}{c}
		h_{1} \\
		h_{2} \\
		h_{3}
	\end{array}
	\right)
	= 
	\left(
	\begin{array}{ccc}
		\sin \beta \sin \gamma & \cos \gamma & 0
		\\
		\sin \beta \cos \gamma & - \sin \gamma & 0
		\\
		\cos \beta &  0 & 1
	\end{array}
	\right)
	\left(
	\begin{array}{c}
		\dot{\alpha} \\
		\dot{\beta} \\
		\dot{\gamma}
	\end{array}
	\right)
\end{eqnarray}
where $H = \sum_{i = 1}^{3} h_{i} \sigma_{i}$ subject to the KAK decomposition given by
\begin{eqnarray} \label{eq:euler}
\exp(tH) = \exp(\alpha(t) \sigma_{3}) \exp(\beta(t) \sigma_{1}) \exp(\gamma(t) \sigma_{3}).
\end{eqnarray}
The KAK decomposition in \eref{eq:euler} together with \eref{eq:eulerangluar} renders the group-theoretic method to obtain the conventional Euler angles $\alpha, \beta, \gamma$ and their derivatives related to the angular velocity discussed in the standard literature such as \cite{goldstein2002}. 

For the cases having a higher rank than $\mathfrak{su}(2)$, one should anticipate having a rather lengthy form for $\bm{J}_{n}$, but an explicit construction of $\bm{J}_{n}$ is always accessible by employing the matrix representation of the adjoint action. In \ref{sec:su4-control}, we illustrate the construction of $\bm{J}_{2}$ via the matrix representation of the adjoint action for $\mathfrak{su}(4)$ although the construction is relatively straightforward in general.
 
\section{Conclusion and outlook}
In this work, we have provided a uniform finite generation of $\textrm{SU}(2^{n})$, which can be taken as an abstract unitary control of a Hamiltonian system embedded in an $n$-qubit system. The uniform finite generation has been obtained by successively applying a decomposition of the KAK type, which has been discussed in \cite{khaneja2001,bullock2004a,dalessandro2007,hai-rui2008}. The decomposition enables one to parameterize $\textrm{SU}(2^{n})$ via angular parameters taken as a generalization of Euler angles for $\textrm{SU}(2)$.

Having the decomposition, we have constructed a system of differential-algebraic equations for the parametric curve in the space of generalized Euler angles, corresponding to the integral curve generated by a given Hamiltonian via the exponential map. The cyclicity of generators of one-parameter subgroups constituting the decomposition allows one to cast DAEs into the form of a matrix equation in \eref{eq:dae-matrix-form} whose matrix consists of trigonometric functions on generalized Euler angles.

As a closing remark, we address two aspects of our result, which may require further investigation concerning the random matrix ensemble and the solvability of DAEs. Firstly, we note that the parameterization in \Sref{sec:KAK} can be taken as an explicit construction of a probabilistic ensemble of $\textrm{SU}(2^{n})$ in the context of random matrix theory. Indeed, the matrix $\bm{J}_{n}$ in \eref{eq:dae-matrix-form} can be identified with the jacobian of the tangent space of $\textrm{SU}(2^{n})$. Therefore, it is straightforward to construct the corresponding volume form (or the probabilistic distribution measure). An explicit representation of such random ensemble has taken a prominent role in quantum machine learning (see for a survey \cite{cerezo2022}), so one may find an application of generalized Euler angles in the context of quantum machine learning. 

Finally, it is worth noting that the conversion of DAEs in \eref{eq:odegeneral} into a system of first-order differential ordinary differential equations is not trivial due to singularities of the decomposition in \eref{eq:kak}, particularly at the identity of $\textrm{SU}(2^{n})$. This singularity issue is closely tied to the solvability of \eref{eq:odegeneral}, which is necessary to justify any attempt to find the integral curve by solving \eref{eq:odegeneral}. In turn, as a future endeavor, a more in-depth investigation into the structure of DAEs (see, e.g., \cite{campbell2006}) through some case studies focused on specific Hamiltonians could be conducted to shed more light on the solvability while expanding the scope and applicability of our method.

\ack

SL is grateful to Joonsuk Huh and Jinhyoung Lee for enlightening discussions on related topics. This work was partly supported by Institute for Information \& Communications Technology Promotion (IITP) grant (No. 2019-0-00003, Research and Development of Core Technologies for Programming, Running, Implementing and Validating of Fault-Tolerant Quantum Computing System), National Research Foundation of Korea (NRF) grant (No.  -2019M3E4A1080146, NRF-2021M3E4A1038213, and NRF-2022M3E4A1077094) and Electronics and Communications Research Institute (ETRI) (23ZB1300, Proprietary Basic Research on Computing Technology for the Disruptive Innovation of Computational Performance) funded by the Korea government (MSIT).

\appendix

\section{A simple case study: $\mathfrak{su}(4)$}
\label{sec:su4-control}

As indicated in \Sref{sec:control}, the construction of $\bm{J}_{n}$ can be implemented by explicitly representing the adjoint action in a matrix form. For $\mathfrak{su}(4)$, by choosing a basis in the following order
\begin{eqnarray*}
    \left( \sigma_{01}, \sigma_{02}, \sigma_{03}, \sigma_{10}, \sigma_{11}, \sigma_{12}, \sigma_{13}, \sigma_{20}, \sigma_{21}, \sigma_{22}, \sigma_{23}, \sigma_{30}, \sigma_{31}, \sigma_{32}, \sigma_{33} \right)
\end{eqnarray*}
one has the matrix representation as
\begin{eqnarray*}
	e^{\textrm{\scriptsize ad}\, \theta\sigma_{10}}\nonumber\\ \simeq
	{\tiny
	\left(\begin{array}{ccccccccccccccc}
		1 & 0 & 0 & 0 & 0 & 0 & 0 & 0 & 0 & 0 & 0 & 0 & 0 & 0 & 0 \\
		0 & 1 & 0 & 0 & 0 & 0 & 0 & 0 & 0 & 0 & 0 & 0 & 0 & 0 & 0 \\
		0 & 0 & 1 & 0 & 0 & 0 & 0 & 0 & 0 & 0 & 0 & 0 & 0 & 0 & 0 \\
		0 & 0 & 0 & 1 & 0 & 0 & 0 & 0 & 0 & 0 & 0 & 0 & 0 & 0 & 0 \\
		0 & 0 & 0 & 0 & 1 & 0 & 0 & 0 & 0 & 0 & 0 & 0 & 0 & 0 & 0 \\
		0 & 0 & 0 & 0 & 0 & 1 & 0 & 0 & 0 & 0 & 0 & 0 & 0 & 0 & 0 \\
		0 & 0 & 0 & 0 & 0 & 0 & 1 & 0 & 0 & 0 & 0 & 0 & 0 & 0 & 0 \\
		0 & 0 & 0 & 0 & 0 & 0 & 0 & \cos \theta & 0 & 0 & 0 & \sin \theta & 0 & 0 & 0 \\
		0 & 0 & 0 & 0 & 0 & 0 & 0 & 0 & \cos \theta & 0 & 0 & 0 & \sin \theta & 0 & 0 \\
		0 & 0 & 0 & 0 & 0 & 0 & 0 & 0 & 0 & \cos \theta & 0 & 0 & 0 & \sin \theta & 0 \\
		0 & 0 & 0 & 0 & 0 & 0 & 0 & 0 & 0 & 0 & \cos \theta & 0 & 0 & 0 & \sin \theta \\
		0 & 0 & 0 & 0 & 0 & 0 & 0 & -\sin \theta & 0 & 0 & 0 & \cos \theta & 0 & 0 & 0 \\
		0 & 0 & 0 & 0 & 0 & 0 & 0 & 0 & -\sin \theta & 0 & 0 & 0 & \cos \theta & 0 & 0 \\
		0 & 0 & 0 & 0 & 0 & 0 & 0 & 0 & 0 & -\sin \theta & 0 & 0 & 0 & \cos \theta & 0 \\
		0 & 0 & 0 & 0 & 0 & 0 & 0 & 0 & 0 & 0 & -\sin \theta & 0 & 0 & 0 & \cos \theta
		\end{array}\right),
	}
	\end{eqnarray*}
	\begin{eqnarray*}	
	e^{\textrm{\scriptsize ad}\, \theta\sigma_{13}}\\ \simeq
	{\tiny
    \left(\begin{array}{ccccccccccccccc}
		\cos \theta & 0 & 0 & 0 & 0 & \sin \theta & 0 & 0 & 0 & 0 & 0 & 0 & 0 & 0 & 0 \\
		0 & \cos \theta & 0 & 0 & -\sin \theta & 0 & 0 & 0 & 0 & 0 & 0 & 0 & 0 & 0 & 0 \\
		0 & 0 & 1 & 0 & 0 & 0 & 0 & 0 & 0 & 0 & 0 & 0 & 0 & 0 & 0 \\
		0 & 0 & 0 & 1 & 0 & 0 & 0 & 0 & 0 & 0 & 0 & 0 & 0 & 0 & 0 \\
		0 & \sin \theta & 0 & 0 & \cos \theta & 0 & 0 & 0 & 0 & 0 & 0 & 0 & 0 & 0 & 0 \\
		-\sin \theta & 0 & 0 & 0 & 0 & \cos \theta & 0 & 0 & 0 & 0 & 0 & 0 & 0 & 0 & 0 \\
		0 & 0 & 0 & 0 & 0 & 0 & 1 & 0 & 0 & 0 & 0 & 0 & 0 & 0 & 0 \\
		0 & 0 & 0 & 0 & 0 & 0 & 0 & \cos \theta & 0 & 0 & 0 & 0 & 0 & 0 & \sin \theta \\
		0 & 0 & 0 & 0 & 0 & 0 & 0 & 0 & 1 & 0 & 0 & 0 & 0 & 0 & 0 \\
		0 & 0 & 0 & 0 & 0 & 0 & 0 & 0 & 0 & 1 & 0 & 0 & 0 & 0 & 0 \\
		0 & 0 & 0 & 0 & 0 & 0 & 0 & 0 & 0 & 0 & \cos \theta & \sin \theta & 0 & 0 & 0 \\
		0 & 0 & 0 & 0 & 0 & 0 & 0 & 0 & 0 & 0 & -\sin \theta & \cos \theta & 0 & 0 & 0 \\
		0 & 0 & 0 & 0 & 0 & 0 & 0 & 0 & 0 & 0 & 0 & 0 & 1 & 0 & 0 \\
		0 & 0 & 0 & 0 & 0 & 0 & 0 & 0 & 0 & 0 & 0 & 0 & 0 & 1 & 0 \\
		0 & 0 & 0 & 0 & 0 & 0 & 0 & -\sin \theta & 0 & 0 & 0 & 0 & 0 & 0 & \cos \theta
		\end{array}\right),
    }
	\end{eqnarray*}
	\begin{eqnarray*}
    e^{\textrm{\scriptsize ad}\,\theta \sigma_{03}}\\ \simeq
    {\tiny
    \left(
		\begin{array}{ccccccccccccccc}
		\cos \theta & \sin \theta & 0 & 0 & 0 & 0 & 0 & 0 & 0 & 0 & 0 & 0 & 0 & 0 & 0 \\
		-\sin \theta & \cos \theta & 0 & 0 & 0 & 0 & 0 & 0 & 0 & 0 & 0 & 0 & 0 & 0 & 0 \\
		0 & 0 & 1 & 0 & 0 & 0 & 0 & 0 & 0 & 0 & 0 & 0 & 0 & 0 & 0 \\
		0 & 0 & 0 & 1 & 0 & 0 & 0 & 0 & 0 & 0 & 0 & 0 & 0 & 0 & 0 \\
		0 & 0 & 0 & 0 & \cos \theta & \sin \theta & 0 & 0 & 0 & 0 & 0 & 0 & 0 & 0 & 0 \\
		0 & 0 & 0 & 0 & -\sin \theta & \cos \theta & 0 & 0 & 0 & 0 & 0 & 0 & 0 & 0 & 0 \\
		0 & 0 & 0 & 0 & 0 & 0 & 1 & 0 & 0 & 0 & 0 & 0 & 0 & 0 & 0 \\
		0 & 0 & 0 & 0 & 0 & 0 & 0 & 1 & 0 & 0 & 0 & 0 & 0 & 0 & 0 \\
		0 & 0 & 0 & 0 & 0 & 0 & 0 & 0 & \cos \theta & \sin \theta & 0 & 0 & 0 & 0 & 0 \\
		0 & 0 & 0 & 0 & 0 & 0 & 0 & 0 & -\sin \theta & \cos \theta & 0 & 0 & 0 & 0 & 0 \\
		0 & 0 & 0 & 0 & 0 & 0 & 0 & 0 & 0 & 0 & 1 & 0 & 0 & 0 & 0 \\
		0 & 0 & 0 & 0 & 0 & 0 & 0 & 0 & 0 & 0 & 0 & 1 & 0 & 0 & 0 \\
		0 & 0 & 0 & 0 & 0 & 0 & 0 & 0 & 0 & 0 & 0 & 0 & \cos \theta & \sin \theta & 0 \\
		0 & 0 & 0 & 0 & 0 & 0 & 0 & 0 & 0 & 0 & 0 & 0 & -\sin \theta & \cos \theta & 0 \\
		0 & 0 & 0 & 0 & 0 & 0 & 0 & 0 & 0 & 0 & 0 & 0 & 0 & 0 & 1
		\end{array}
		\right),
    }
    \end{eqnarray*}
	\begin{eqnarray*}
    e^{\textrm{\scriptsize ad}\,\theta \sigma_{01}}\\ \simeq 
    {\tiny 
		\left(\begin{array}{ccccccccccccccc}
		1 & 0 & 0 & 0 & 0 & 0 & 0 & 0 & 0 & 0 & 0 & 0 & 0 & 0 & 0 \\
		0 & \cos \theta & \sin \theta & 0 & 0 & 0 & 0 & 0 & 0 & 0 & 0 & 0 & 0 & 0 & 0 \\
		0 & -\sin \theta & \cos \theta & 0 & 0 & 0 & 0 & 0 & 0 & 0 & 0 & 0 & 0 & 0 & 0 \\
		0 & 0 & 0 & 1 & 0 & 0 & 0 & 0 & 0 & 0 & 0 & 0 & 0 & 0 & 0 \\
		0 & 0 & 0 & 0 & 1 & 0 & 0 & 0 & 0 & 0 & 0 & 0 & 0 & 0 & 0 \\
		0 & 0 & 0 & 0 & 0 & \cos \theta & \sin \theta & 0 & 0 & 0 & 0 & 0 & 0 & 0 & 0 \\
		0 & 0 & 0 & 0 & 0 & -\sin \theta & \cos \theta & 0 & 0 & 0 & 0 & 0 & 0 & 0 & 0 \\
		0 & 0 & 0 & 0 & 0 & 0 & 0 & 1 & 0 & 0 & 0 & 0 & 0 & 0 & 0 \\
		0 & 0 & 0 & 0 & 0 & 0 & 0 & 0 & 1 & 0 & 0 & 0 & 0 & 0 & 0 \\
		0 & 0 & 0 & 0 & 0 & 0 & 0 & 0 & 0 & \cos \theta & \sin \theta & 0 & 0 & 0 & 0 \\
		0 & 0 & 0 & 0 & 0 & 0 & 0 & 0 & 0 & -\sin \theta & \cos \theta & 0 & 0 & 0 & 0 \\
		0 & 0 & 0 & 0 & 0 & 0 & 0 & 0 & 0 & 0 & 0 & 1 & 0 & 0 & 0 \\
		0 & 0 & 0 & 0 & 0 & 0 & 0 & 0 & 0 & 0 & 0 & 0 & 1 & 0 & 0 \\
		0 & 0 & 0 & 0 & 0 & 0 & 0 & 0 & 0 & 0 & 0 & 0 & 0 & \cos \theta & \sin \theta \\
		0 & 0 & 0 & 0 & 0 & 0 & 0 & 0 & 0 & 0 & 0 & 0 & 0 & -\sin \theta & \cos \theta
		\end{array}
		\right),
    }
\end{eqnarray*}

\begin{eqnarray*}
    e^{\textrm{\scriptsize ad}\, \theta \sigma_{30}}\\ \simeq 
    {\tiny
	\left(\begin{array}{ccccccccccccccc}
		1 & 0 & 0 & 0 & 0 & 0 & 0 & 0 & 0 & 0 & 0 & 0 & 0 & 0 & 0 \\
		0 & 1 & 0 & 0 & 0 & 0 & 0 & 0 & 0 & 0 & 0 & 0 & 0 & 0 & 0 \\
		0 & 0 & 1 & 0 & 0 & 0 & 0 & 0 & 0 & 0 & 0 & 0 & 0 & 0 & 0 \\
		0 & 0 & 0 & \cos \theta & 0 & 0 & 0 & \sin \theta & 0 & 0 & 0 & 0 & 0 & 0 & 0 \\
		0 & 0 & 0 & 0 & \cos \theta & 0 & 0 & 0 & \sin \theta & 0 & 0 & 0 & 0 & 0 & 0 \\
		0 & 0 & 0 & 0 & 0 & \cos \theta & 0 & 0 & 0 & \sin \theta & 0 & 0 & 0 & 0 & 0 \\
		0 & 0 & 0 & 0 & 0 & 0 & \cos \theta & 0 & 0 & 0 & \sin \theta & 0 & 0 & 0 & 0 \\
		0 & 0 & 0 & -\sin \theta & 0 & 0 & 0 & \cos \theta & 0 & 0 & 0 & 0 & 0 & 0 & 0 \\
		0 & 0 & 0 & 0 & -\sin \theta & 0 & 0 & 0 & \cos \theta & 0 & 0 & 0 & 0 & 0 & 0 \\
		0 & 0 & 0 & 0 & 0 & -\sin \theta & 0 & 0 & 0 & \cos \theta & 0 & 0 & 0 & 0 & 0 \\
		0 & 0 & 0 & 0 & 0 & 0 & -\sin \theta & 0 & 0 & 0 & \cos \theta & 0 & 0 & 0 & 0 \\
		0 & 0 & 0 & 0 & 0 & 0 & 0 & 0 & 0 & 0 & 0 & 1 & 0 & 0 & 0 \\
		0 & 0 & 0 & 0 & 0 & 0 & 0 & 0 & 0 & 0 & 0 & 0 & 1 & 0 & 0 \\
		0 & 0 & 0 & 0 & 0 & 0 & 0 & 0 & 0 & 0 & 0 & 0 & 0 & 1 & 0 \\
		0 & 0 & 0 & 0 & 0 & 0 & 0 & 0 & 0 & 0 & 0 & 0 & 0 & 0 & 1
		\end{array}\right),
    }
	\end{eqnarray*}
	\begin{eqnarray*}
    e^{\textrm{\scriptsize ad}\, \theta \sigma_{33}}\\ \simeq
    {\tiny
		\left(\begin{array}{ccccccccccccccc}
		\cos \theta & 0 & 0 & 0 & 0 & 0 & 0 & 0 & 0 & 0 & 0 & 0 & 0 & \sin \theta & 0 \\
		0 & \cos \theta & 0 & 0 & 0 & 0 & 0 & 0 & 0 & 0 & 0 & 0 & -\sin \theta & 0 & 0 \\
		0 & 0 & 1 & 0 & 0 & 0 & 0 & 0 & 0 & 0 & 0 & 0 & 0 & 0 & 0 \\
		0 & 0 & 0 & \cos \theta & 0 & 0 & 0 & 0 & 0 & 0 & \sin \theta & 0 & 0 & 0 & 0 \\
		0 & 0 & 0 & 0 & 1 & 0 & 0 & 0 & 0 & 0 & 0 & 0 & 0 & 0 & 0 \\
		0 & 0 & 0 & 0 & 0 & 1 & 0 & 0 & 0 & 0 & 0 & 0 & 0 & 0 & 0 \\
		0 & 0 & 0 & 0 & 0 & 0 & \cos \theta & \sin \theta & 0 & 0 & 0 & 0 & 0 & 0 & 0 \\
		0 & 0 & 0 & 0 & 0 & 0 & -\sin \theta & \cos \theta & 0 & 0 & 0 & 0 & 0 & 0 & 0 \\
		0 & 0 & 0 & 0 & 0 & 0 & 0 & 0 & 1 & 0 & 0 & 0 & 0 & 0 & 0 \\
		0 & 0 & 0 & 0 & 0 & 0 & 0 & 0 & 0 & 1 & 0 & 0 & 0 & 0 & 0 \\
		0 & 0 & 0 & -\sin \theta & 0 & 0 & 0 & 0 & 0 & 0 & \cos \theta & 0 & 0 & 0 & 0 \\
		0 & 0 & 0 & 0 & 0 & 0 & 0 & 0 & 0 & 0 & 0 & 1 & 0 & 0 & 0 \\
		0 & \sin \theta & 0 & 0 & 0 & 0 & 0 & 0 & 0 & 0 & 0 & 0 & \cos \theta & 0 & 0 \\
		-\sin \theta & 0 & 0 & 0 & 0 & 0 & 0 & 0 & 0 & 0 & 0 & 0 & 0 & \cos \theta & 0 \\
		0 & 0 & 0 & 0 & 0 & 0 & 0 & 0 & 0 & 0 & 0 & 0 & 0 & 0 & 1
		\end{array}\right).
    }
\end{eqnarray*}

Then by employing
\begin{eqnarray*}
    \Pi_{03}(x) =
    {\tiny 
	\left(
    \begin{array}{c}
        0\\
        0\\
        x\\
        0\\
        0\\
        0\\
        0\\
        0\\
        0\\
        0\\
        0\\
        0\\
        0\\
        0\\
        0
    \end{array}
	\right)
    },\quad
    \Pi_{01}(x) =
    {\tiny 
	\left(
    \begin{array}{c}
        x\\
        0\\
        0\\
        0\\
        0\\
        0\\
        0\\
        0\\
        0\\
        0\\
        0\\
        0\\
        0\\
        0\\
        0
    \end{array}
	\right)
    },\quad
    \Pi_{30}(x) =
    {\tiny 	
	\left(
    \begin{array}{c}
        0\\
        0\\
        0\\
        0\\
        0\\
        0\\
        0\\
        0\\
        0\\
        0\\
        0\\
        x\\
        0\\
        0\\
        0
    \end{array}
	\right)
    },\\
    \Pi_{33}(x) =
    {\tiny
	\left( 
    \begin{array}{c}
        0\\
        0\\
        0\\
        0\\
        0\\
        0\\
        0\\
        0\\
        0\\
        0\\
        0\\
        0\\
        0\\
        0\\
        x
    \end{array}
	\right)
    },\quad
    \Pi_{10}(x) =
	\left(
    {\tiny 
    \begin{array}{c}
        0\\
        0\\
        0\\
        x\\
        0\\
        0\\
        0\\
        0\\
        0\\
        0\\
        0\\
        0\\
        0\\
        0\\
        0
    \end{array}
	}\right),\quad
    \Pi_{03}(x) =
	\left(
    {\tiny 
    \begin{array}{c}
        0\\
        0\\
        0\\
        0\\
        0\\
        x\\
        0\\
        0\\
        0\\
        0\\
        0\\
        0\\
        0\\
        0
    \end{array}
	} \right),
\end{eqnarray*}
one obtains DAEs for $\mathfrak{su}(4)$ as
\begin{eqnarray*}
    \bm{h} = 
    &
    \prod_{i = 18}^{2} e^{- \textrm{\scriptsize ad}\, \theta_{i} X_{i}} \Pi_{03}(\dot{\theta}_{1}) 
    +
    \prod_{i = 18}^{3} e^{- \textrm{\scriptsize ad}\, \theta_{i} X_{i}} \Pi_{01}(\dot{\theta}_{2})
	+
    \prod_{i = 18}^{4} e^{- \textrm{\scriptsize ad}\, \theta_{i} X_{i}} \Pi_{03}(\dot{\theta}_{3})
    \\
	&
	+
    \prod_{i = 18}^{5} e^{- \textrm{\scriptsize ad}\, \theta_{i} X_{i}} \Pi_{30}(\dot{\theta}_{4})
	+
    \prod_{i = 18}^{6} e^{- \textrm{\scriptsize ad}\, \theta_{i} X_{i}} \Pi_{33}(\dot{\theta}_{5})
    +
    \prod_{i = 18}^{7} e^{- \textrm{\scriptsize ad}\, \theta_{i} X_{i}} \Pi_{03}(\dot{\theta}_{6})
    \\
	&
	+
    \prod_{i = 18}^{8} e^{- \textrm{\scriptsize ad}\, \theta_{i} X_{i}} \Pi_{01}(\dot{\theta}_{7})
    +
    \prod_{i = 18}^{9} e^{- \textrm{\scriptsize ad}\, \theta_{i} X_{i}} \Pi_{03}(\dot{\theta}_{8})
    +
    \prod_{i = 18}^{10} e^{- \textrm{\scriptsize ad}\, \theta_{i} X_{i}} \Pi_{10}(\dot{\theta}_{9})
    \\
	&
	+
    \prod_{i = 18}^{11} e^{- \textrm{\scriptsize ad}\, \theta_{i} X_{i}} \Pi_{13}(\dot{\theta}_{10})
    +
    \prod_{i = 18}^{12} e^{- \textrm{\scriptsize ad}\, \theta_{i} X_{i}} \Pi_{03}(\dot{\theta}_{11})
    +
    \prod_{i = 18}^{13} e^{- \textrm{\scriptsize ad}\, \theta_{i} X_{i}} \Pi_{01}(\dot{\theta}_{12})
    \\
	&
	+
    \prod_{i = 18}^{14} e^{- \textrm{\scriptsize ad}\, \theta_{i} X_{i}} \Pi_{03}(\dot{\theta}_{13})
    +
    \prod_{i = 18}^{15} e^{- \textrm{\scriptsize ad}\, \theta_{i} X_{i}} \Pi_{30}(\dot{\theta}_{14})
	+
    \prod_{i = 18}^{16} e^{- \textrm{\scriptsize ad}\, \theta_{i} X_{i}} \Pi_{33}(\dot{\theta}_{15})
    \\
	&
	+
    \prod_{i = 18}^{17} e^{- \textrm{\scriptsize ad}\, \theta_{i} X_{i}} \Pi_{03}(\dot{\theta}_{16})
	+ 
    e^{- \textrm{\scriptsize ad}\, \theta_{18} X_{18}} \Pi_{01}(\dot{\theta}_{17})
    +
    \Pi_{03}(\dot{\theta}_{18})
\end{eqnarray*}
together with $\bm{h} = \left( h_{1}, h_{2}, \ldots, h_{15} \right)^{T}$.

\section*{References}
\bibliographystyle{iopart-num.bst}

\providecommand{\newblock}{}

\end{document}